\documentclass[a4paper]{spie}  
\usepackage{amsmath,amsfonts,amssymb}
\usepackage{graphicx}
\usepackage[colorlinks=true, allcolors=blue]{hyperref}

\usepackage{booktabs}

\newcommand{\dw}{$\lambda/D$}

\title{High contrast imaging for the enhanced resolution imager and spectrometer (ERIS)}

\author[a]{Matthew A. Kenworthy}
\author[a]{Frans Snik}
\author[a]{Christoph U. Keller}
\author[a]{David Doelman}
\author[a]{Emiel H. Por}
\author[b]{Olivier Absil}
\author[b]{Brunella Carlomagno}
\author[c]{Mikael Karlsson}
\author[d,e]{Elsa Huby}
\author[f]{Adrian M. Glauser}
\author[f]{Sascha P. Quanz}
\author[g]{William D. Taylor}

\affil[a]{Leiden Observatory, Leiden University, P.O. Box 9513, 2300 RA Leiden, The Netherlands}
\affil[b]{Space sciences, Technologies, and Astrophysics Research (STAR) Institute, Universit{\'e} de Li{\`e}ge, 19c all{\'e}e du Six Ao{\^u}t, B-4000 Sart Tilman, Belgium}
\affil[c]{Angstro{\"m} Laboratory, Uppsala University, L{\"a}gerhyddsv{\"a}gen 1, SE-751 21 Uppsala, Sweden}
\affil[d]{Lab. d'Etudes Spatiales et d'Instrumentation en Astrophysique (France)}
\affil[e]{Observatoire de Paris {\`a} Meudon (France)}
\affil[f]{Institute for Astronomy, ETH Z{\"u}rich, Wolfgang-Pauli-Str. 27, 8093 Z{\"u}rich, Switzerland}
\affil[g]{ UK Astronomy Technology Centre, STFC, Blackford Hill, Edinburgh, EH9 3HJ, UK}

\authorinfo{E-mail: \href{mailto:kenworthy@strw.leidenuniv.nl}{kenworthy@strw.leidenuniv.nl}, Telephone: +31 (0) 71 527 8455}

\begin{document} 
\maketitle

\begin{abstract}
ERIS is a diffraction limited thermal infrared imager and spectrograph for the Very Large Telescope UT4.
One of the science cases for ERIS is the detection and characterization of circumstellar structures and exoplanets around bright stars that are typically much fainter than the stellar diffraction halo.
Enhanced sensitivity is provided through the combination of (i) suppression of the diffraction halo of the target star using coronagraphs, and (ii) removal of any residual diffraction structure through focal plane wavefront sensing and subsequent active correction.
%
%
In this paper we present the two coronagraphs used for diffraction suppression and enabling high contrast imaging in ERIS.

\end{abstract}

\keywords{ERIS, NIX, VLT, Adaptive Optics, Apodizing Phase Plate, Vortex Coronagraph, Focal Plane Coronagraphy}

\section{INTRODUCTION}
\label{sec:intro}  

ERIS is a next generation imager for the Very Large Telescope (VLT) with an expected first light of 2020 at the Cassegrain focus of UT4, and the general overview of the instrument and science cases is detailed in these proceedings.
The detection of extrasolar planets (exoplanets) around stars in our Galaxy has revolutionised our view of planet formation, evolution and the architecture of planetary systems.
Many of these exoplanet detections are indirect, in that they are discovered by their effect on their parent star.
The radial velocity method, where the reflex motion of the star around the barycentre of the star/planet system is measured spectroscopically, has detected many hundreds of planets.
When the orbit of the exoplanet has an inclination close to 90 degrees, the planet will transit in front of the star and be detected by photometric monitoring and for a subset of these transiting systems, spectroscopic measurements reveal the signatures of molecules in their atmospheres.
Both RV and transit methods preferentially detect planets on orbital periods of two decades or less, and in most cases do not provide information on the composition of the exoplanet itself. 
Direct imaging of extrasolar planets explores longer orbital period planets and enables characterization of the atmosphere through both reflected light and direct thermal emission.
Variations in cloud cover are revealed in variability measured across multiple epochs.

Several large direct imaging exoplanet surveys with extreme adaptive optics systems (e.g. SPHERE, GPI, SCExAO) have focused on the short infrared bands from 1 to 2.5 microns.
Exoplanet cooling models show that planets are red in color, and their emergent flux from 3 to 5 microns falls more slowly with age than shorter wavelengths.
In this paper, we describe the High Contrast Imaging (HCI) modes for ERIS, which use the grating vector Apodizing Phase Plate and Vortex Coronagraph to provide enhanced sensitivity to faint companions and structures next to bright stars.
%

\section{Principle and location of the Coronagraphs}

There are two coronagraphs implemented in ERIS, the grating vector Apodizing Phase Plate and the Vortex Coronagraph.
The grating vector Apodizing Phase Plate coronagraph (gvAPP)\cite{Otten14} is a realization of an APP coronagraph\cite{Codona04,Kenworthy07} using vector phase to implement the phase pattern on the pupil of the telescope\cite{Snik12}.
The Vortex Coronagraph (VC) is implemented using two optics - an Annular Groove Phase Mask (AGPM)\cite{Mawet05} in the telescope focal plane and a Lyot Stop (LS) in the pupil plane, and have een implemented on VLT/NACO\cite{Mawet13b} and Keck\cite{Serabyn17}.

\begin{figure}
\centering
\includegraphics[width=\textwidth]{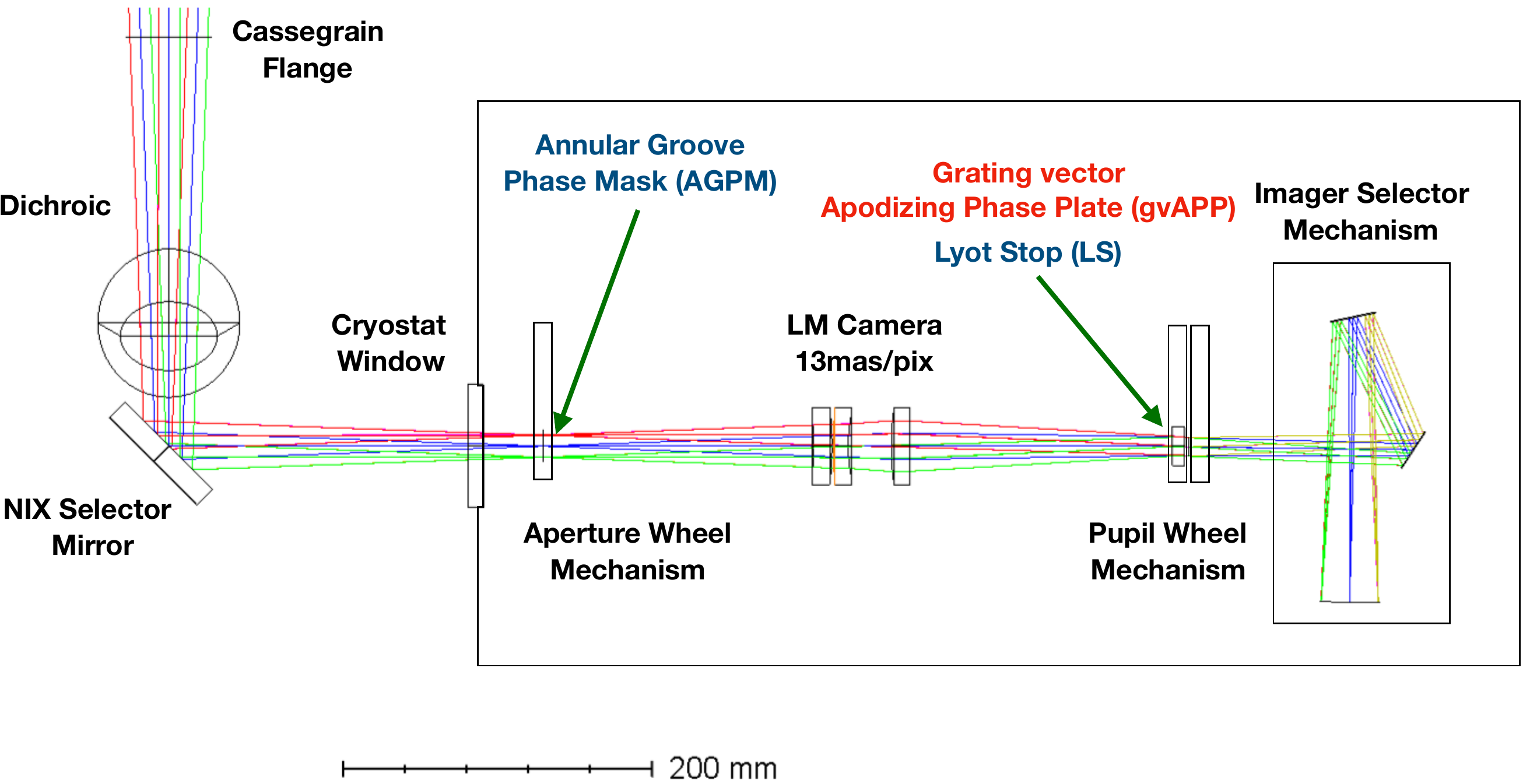}
\caption{Optical layout of ERIS and the locations of the coronagraphic masks.}
\label{fig:erislayout}
\end{figure}

ERIS is composed of an integral field spectrograph SPIFFIER and the NIX imager\cite{Pearson16}.
NIX provides diffraction limited imaging from J band to M band over a field of up to 30 arcseconds with a 13mas pixel scale.
Light from the telescope comes in through the cryostat window and forms a focus coincident with the optical plane of the Aperture Wheel (APW) which can put in various focal plane masks and slits into the beam.
The light then passes through one of three interchangable optical barrels which select for pixel scale and wavelength band.
A second mechanism houses the Pupil and Filter Wheel (PFW), and then passes to the imaging array via several fold mirrors.
The location of the coronagraphic optics are shown in Figure~\ref{fig:erislayout}.

All coronagraphic observations are carried out in pupil tracking mode, which means that for the alt-az Very Large Telescope the sky rotates during the science observations but relative motion between the telescope and camera optics is minimized.
Observations are then analyzed using Angular Differential Imaging in combination with one of several PSF estimation algorithms to form a final science image.

\subsection{The grating vector Apodizing Phase Plate}

\begin{figure}
\centering
\includegraphics[width=\textwidth]{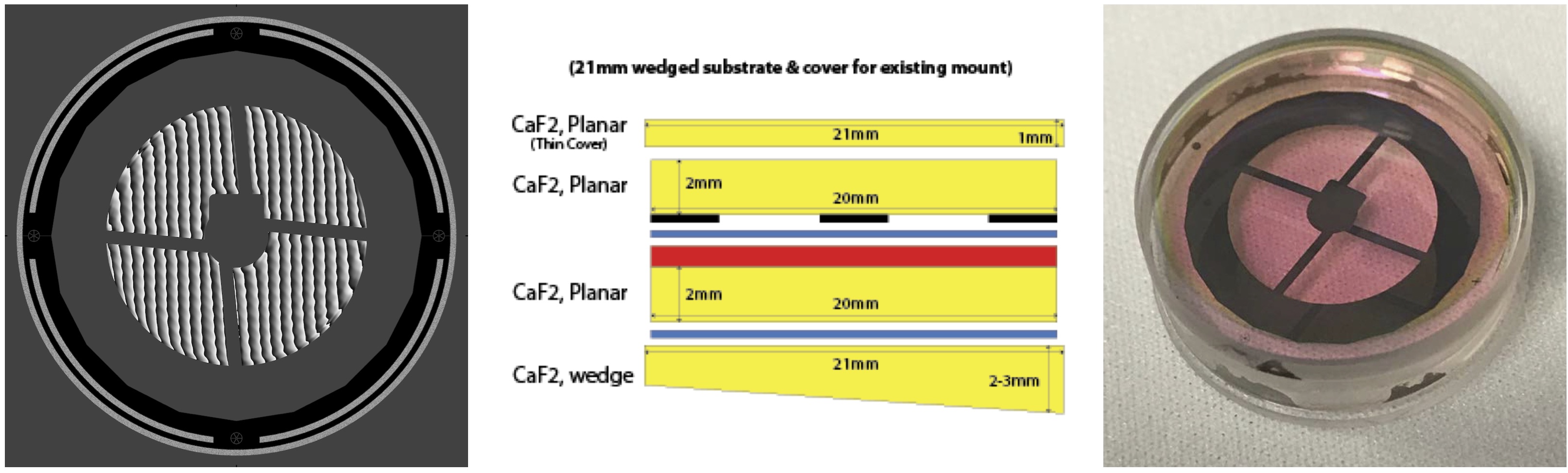}
\caption{Manufacture of the gvAPP. From left to right: The calculated phase pattern and amplitude mask of the gvAPP. The figure shows phase wrapping seen when the optic is held between crossed polarisers. The central panel shows the assembly of the gvAPP, with one layer containing the pupil/ampliutde mask, and the lqiuid crystal layer. The right hand photo shows the assembled and manufactured gvAPP.}
\label{fig:gvapp_manu}
\end{figure}

The gvAPP modifies the telescope PSF by modifying the wavefront in phase only - there is one pupil plane optic located in the PFW.
The resultant PSF for all point sources in the science camera focal plane have a dark D-shaped region from 2 to 7 $\lambda/D$ cleared on one side of the target star.
This PSF modification works on all wavelengths from 2 to 5 microns.
The VLT pupil is circular, and consists of the primary mirror with light blocked by the secondary support arms, the secondary hub itself, and additional obscurations specific to the VLT.
The tertiary mirror that sends light from the secondary mirror across the primary to the Nasmyth foci is stowed out of the Cassegrain focus beam, and the mirror and optomechanical support is folded up so that the short axis of the tertiary mirror obscures the telescope beam from the primary mirror.
This appears as a rectangular obscuration next to the secondary mirror obscuration.
Additionally there is a stray light baffle installed on the end of the secondary hub that prevents scattered light passing by the secondary hub and into any optical wide field imagers that are used at the VLT.
Alignment tolerances and internal flexure within ERIS mean that the pupil image within ERIS will move away from a nominal location with changing telescope elevation.
For correct operation of the gvAPP, a pupil mask is defined at the location of the telescope pupil, and it is undersized such that the gvAPP pupil is guaranteed to be illuminated by the telescope pupil for all orientations and flexure of the telescope and instrument.
This results in a reduction in throughput but ensures optimum operation.
The assembled gvAPP is shown in Figure~\ref{fig:gvapp_manu}.

The optic has a diameter of 20.95mm and a thickness of 7.55mm and is made up of a sandwich of several layers of Calcium Flouride.
Antireflection coatings on the outer surfacese have a transmission $>99.5\%$, and the combined wavefront error over the diameter of the optic is $\lambda/20$ at $3\mu m$.
The total average transmission of the gvAPP optic is greater than 60\%, with a transmission of 95\% at 2 microns, falling to 50\% at 4.5 microns.
A narrow absorption feature due to the NOA-61 glue is present at 3.4 microns.
The pupil mask is made up of aluminium metal layer 300nm thick, resulting in a neutral density greater than 4.
The diameter of the pupil mask is 12.24mm in diameter.
The active optical component of the gvAPP is three layers of liquid crystal polymer totalling 18microns in thickness that together form an achromatic phase retardance from 2 to 5 microns.

\begin{figure}
\centering
\includegraphics[width=0.8\textwidth]{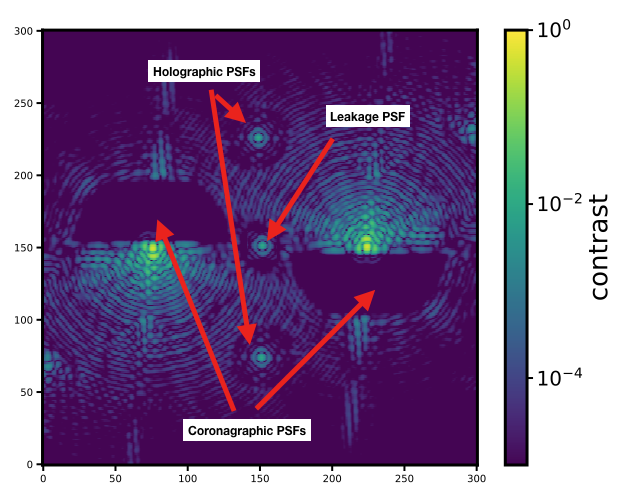}
\caption{The three PSFs formed by the gvAPP on the ERIS detector, and two holographic PSFs that aid in focus calibration.}
\label{fig:gvapp_annotate}
\end{figure}

Three stellar images are formed on the detector - two coronagraphic PSFs with a dark D-shaped hole on one side of the stellar image, and an undeviated image with the direct imaging PSF of the gvAPP pupil but without the coronagraphic PSF, referred to as the `leakage term PSF'.
The majority of the stellar (and planetary) flux is split between the two coronagraphic PSFs, with the remaining 2 to 5\% of stellar flux in the leakage term PSF (see Figure~\ref{fig:gvapp_annotate}).
The gvAPP works optimally with narrow band filters in the L and M bands, and with limited efficiency in the K band.
Using broad band filters will result in chromatic smearing of the the stellar and companion PSF into a low spectral resolution spectrum.

Observing modes for the gvAPP will utilize beam switching on the detector, the baseline observation mode will be to beamswitch the target star between two sides of the imaging array on a duty cycle of 30 seconds and modified by on-sky observational conditions.
The gvAPP is ideally suited for astrometric measurements of companions, and for photometric monitoring.
The individual exposure times for the gvAPP are determined by the integration time required for the flux measured in the D-shaped coronagraphic holes to reach 70\% of the full well depth of the detector.
In the case of bright stars, this signal in the dark hole is expected to be dominated by the flux from the adaptive optics corrected PSF.
This typically results in saturation of the Airy cores of the coronagraphic PSFs, which makes them useless for precision differential photometry.
The leakage term PSF will not be typically saturated on the detector as it is a few percent of the incident stellar flux.
As it is generated through the same optical train through ERIS, it therefore becomes the ideal photometric and astrometric reference for any secondary companions detected in the coronagraphic PSFs.

\subsection{Vortex Coronagraph}

\begin{figure}
\centering
\includegraphics[width=\textwidth]{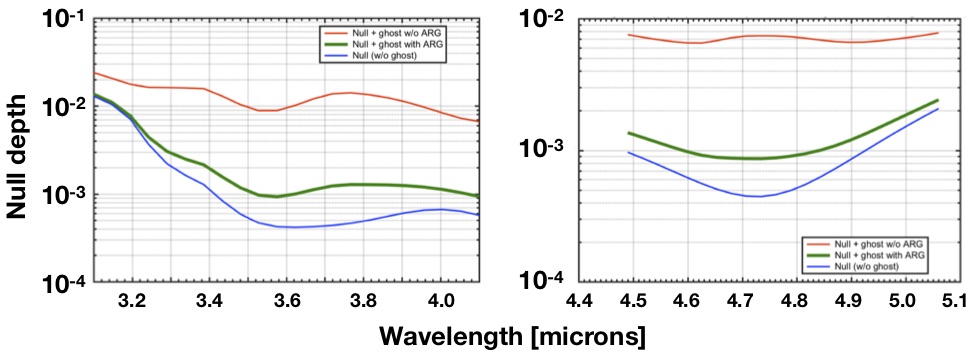}
\caption{Calculated null depth for L and M band AGPMs. Systematic effects added include multiple internal reflections and the effect of an antireflection structure etched into the flat side of the AGPM.}
\label{fig:nulls}
\end{figure}

The Vortex Coronagraph (VC) is realized as two separate optics in ERIS.
The first optic is the Annular Groove Phase Mask (AGPM) that is in the focal plane and inserted into the telescope beam in the Aperture Wheel.
The AGPM is realized on a etched diamond\cite{VargasCatalan16} substrate that is 0.3mm thick by 10mm in diameter and imparts a topological charge of 2.
It has sub-wavelength circular grooves etched into one side that cover the whole side of the optic and suppress the on-axis light of a diffraction limited source.
An anti-reflection etching on the unetched side increases the transmission of the optic - the combined effect of wavelength dependence, internal reflections and etching are shown in Figure~\ref{fig:nulls}.
Null depths of $3\times 10^{-3}$ and $10^{-2}$ for L and M band respectively have been measured for the manufactured AGPM optics for ERIS.
The diffraction limited image of a star is placed on the centre of the AGPM, adding a ramp of phase to the telescope beam around the central optical singularity, and the resultant propagation into the far field regime moves the light from the star outside the nominal telescope pupil.

\begin{figure}
\centering
\includegraphics[width=0.8\textwidth]{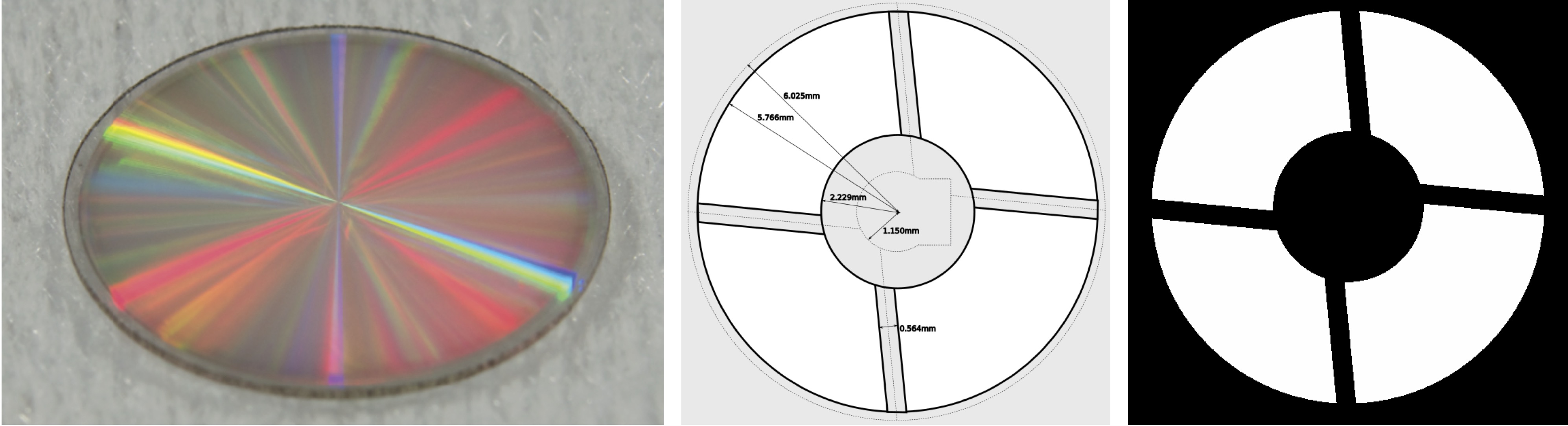}
\caption{On the left, the annular groove phase mask for L band. The middle and right images show the design of the Lyot Stop for the AGPM and vortex coronagraph.}
\label{fig:vortexmanu}
\end{figure}

\begin{figure}
\centering
\includegraphics[width=0.8\textwidth]{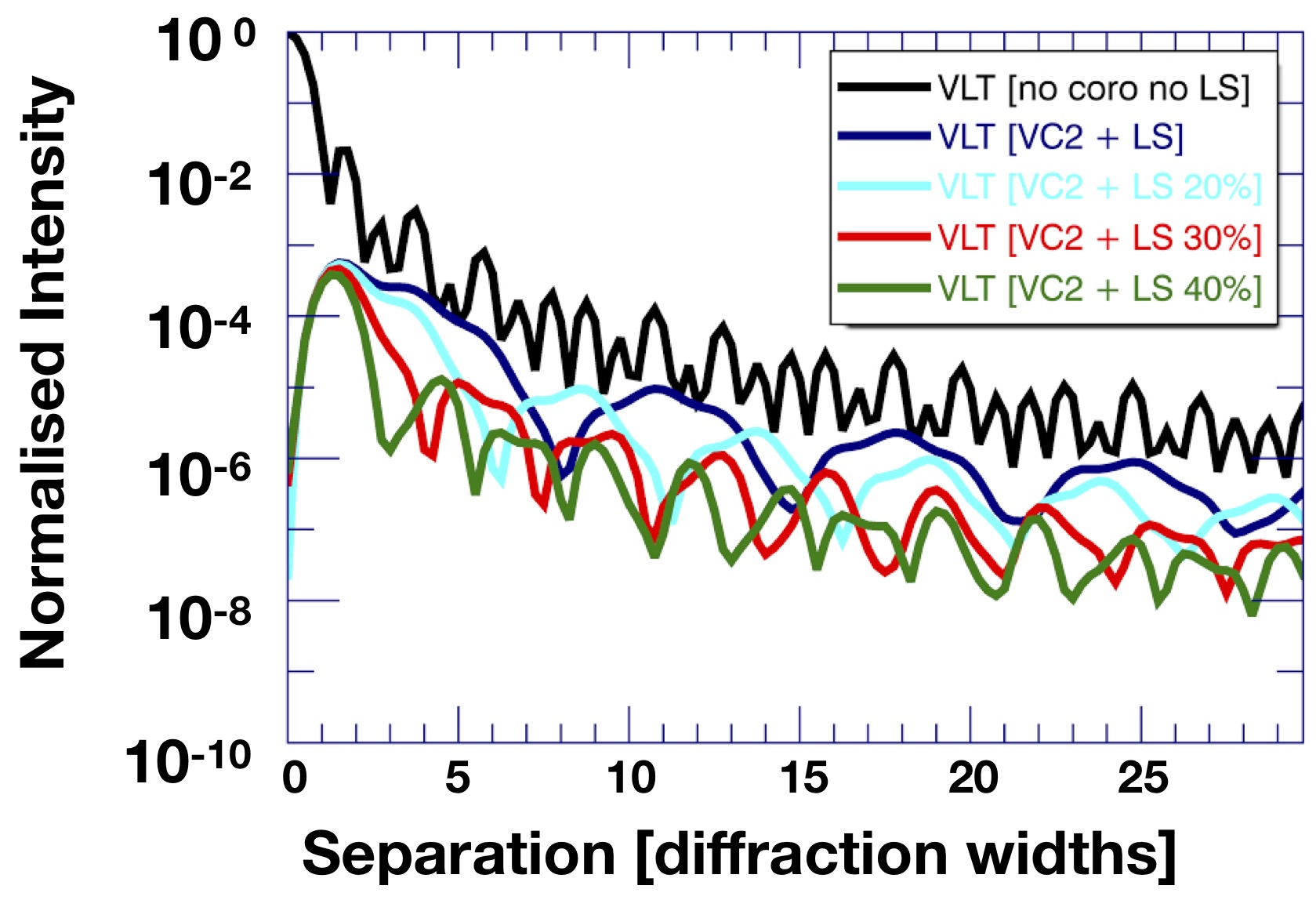}
\caption{Angular separation versus intensity for a Vortex Coronagraph for different central obscuration diameters. The LS for ERIS is set at 30\%.}
\label{fig:vortexsecond}
\end{figure}

Any PSF that is not centered on the central singularity does not have the phase ramp applied to it, and the light distribution in the pupil plane is similar to that of a directly imaged point source.
An ideal circular filled pupil will have all starlight scattered outside the telescope pupil, and using a circular Lyot Stop with the same diameter as the telescope pupil will reject all starlight and transmit all off-axis point sources.
In reality, the secondary hub and the secondary support arms scatter starlight into the telescope pupil, and so a trade off between maximizing the throughput of the off-axis sources and minimizing the transmitted starlight.
For ERIS, this results in a Lyot mask that has a central obscuration diameter of 30\% of the telescope pupil diameter, and oversized secondary support structures for the mask (see Figures~\ref{fig:vortexsecond} and \ref{fig:vortexmanu}).
Separate AGPMs are manufactured for the L and M bands, located in the Aperture Wheel.
One Lyot stop is used for both AGPMs.

The observing strategy for the Vortex Coronagraph is to center the star on the optical axis of the APGM, observe for 10 to 60 seconds depending on the sky background levels, and repeat until 5 minutes is reached.
The star is moved off the detector and a flat fielding exposure is taken, and the observing cycle repeated up to the total on-sky integration time.
The flat field compensates for variations in transmission across the detector and the AGPM optic.

\section{Tip-Tilt and Higher Order Active Correction}

\begin{figure}
\begin{center}
\includegraphics[width=0.80\textwidth]{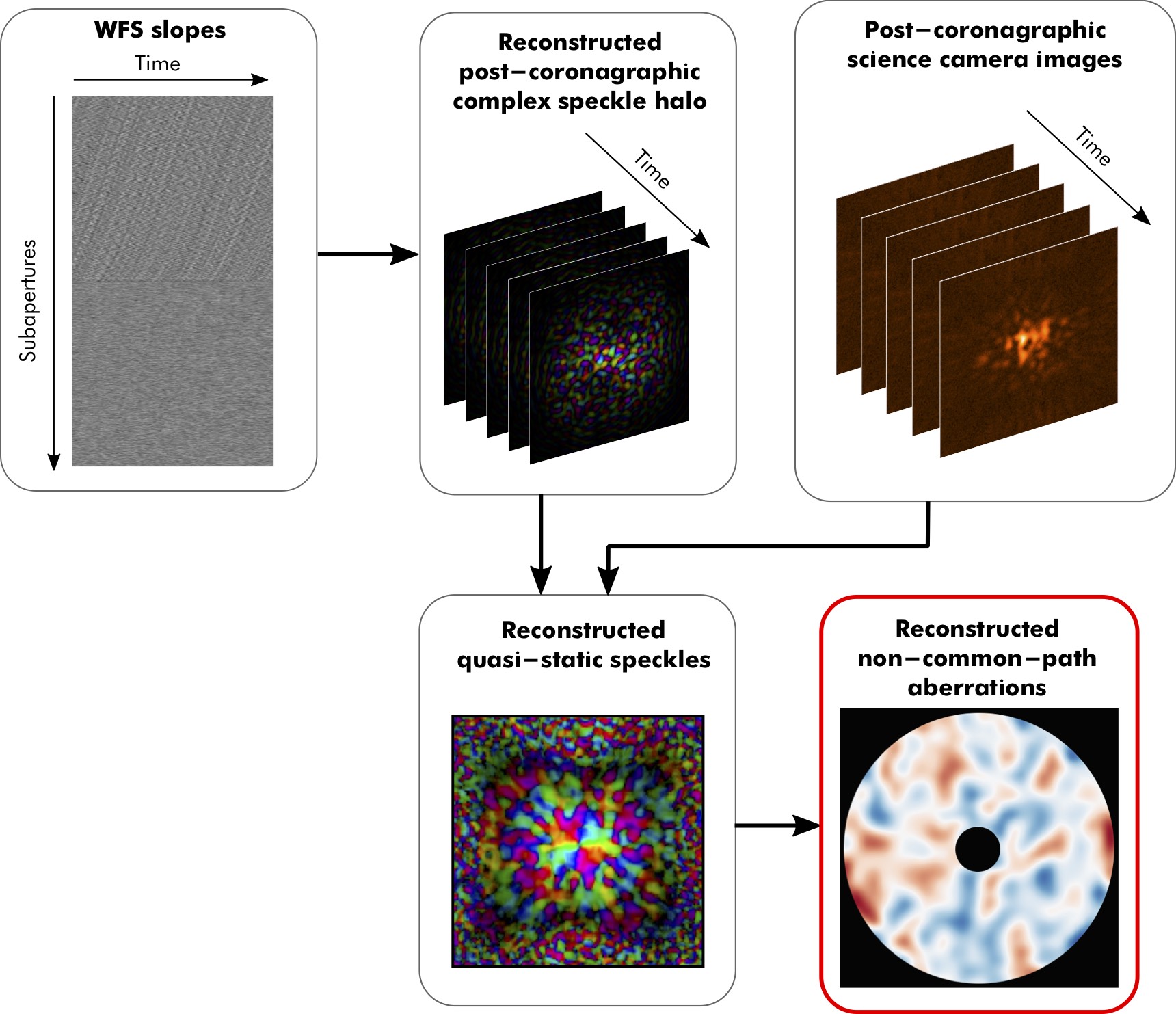}
	\caption{\label{fig:PSI}Schematic showing the process of PSI.}
\end{center}
\end{figure}

The VC is sensitive to both atmospheric dispersion and tip-tilt errors, both from rapid vibrations and from 30 second drifts that build up between the wavefront sensor camera optics and the science camera optics.
To ensure that the star is aligned precisely on the central defect, the science camera frames are analyzed by an algorithm called QACITS\cite{Huby15}.
This algorithm takes the residual donut image that the VC produces on the science camera, and calculates the low order moments of the light distribution around this donut.
A perfectly centered star produces a donut that is evenly illuminated in the azimuthal direction.
When the star moves off the central defect by a fraction of a diffraction width, the light distribution around the donut changes and this will be used as an error signal that will be fed back to the AO control system for ERIS.
An error signal will be applied every 30 seconds to compensate for the long term drift in alignment, with a measurement accuracy of 4mas. 
This system was implemented successfully on the Keck telescope to maintain the performance of the VC\cite{Huby17}.

The science camera focal plane does not sense the same optical path as the wavefront sensor camera does.
This non-common path aberration (NCPA) error results in sub-optimal performance of the coronagraphs.
%

For higher order non-common path aberrations between the science camera focal planes and the wavefront sensor cameras, Phase Sorting Interferometry (PSI)\cite{Codona13} will be used.
Wavefront sensor telemetry is recorded simultaneously with science camera images that have integration times on the order of 10-100 milliseconds.
The telemetry is correlated with the speckles seen in the science camera focal plane from 2 to 7 \dw\ to produce a map of the non-common path aberrations, as shown in Figure~\ref{fig:PSI}.
These time varying aberrations can either be used in post-processing to generate the science camera PSF for all science exposures, or dynamically applied to the deformable mirror as a static wavefront offset in the closed loop operation to reduce the photon shot noise introduced by long lived speckles in the dark region produced by the coronagraphs.

\section{Conclusions}

We have presented the high contrast imaging coronagraphs that are implemented for ERIS, along with the algorithms that will ensure high performance during science observations.
Two coronagraphs will be used: the gvAPP is ideal for discovery, astrometry and photometric variability of point sources, and can carry out beamswitching observations in suboptimal conditions and at higher airmasses.
The VC is ideal for the imaging of circumstellar disks and other extended structures around target stars, and performs extremely well when telescope vibrations and wind shake are minimized.


\acknowledgments 
 
This research made use of {\tt astropy}, a community-developed core Python package for Astronomy \cite{2013A&A...558A..33A}.
%

\bibliography{report} 
\bibliographystyle{spiebib} 

\end{document}